# Hollow-core fiber Characterization with Correlation-Optical Time Domain Reflectometry

**Florian Azendorf[1,2], Bernhard Schmauss[2], Bo Shi[3], Eric Numkam Fokoua[3], Radan Slavík[3], Michael Eiselt[1]**

[1] *ADVA Optical Networking SE, Maerzenquelle 1-3, 98617 Meiningen, Germany*
[2] *LHFT, Friedrich-Alexander-Universität Erlangen-Nürnberg, 91058 Erlangen, Germany*
[3] *Optoelectronics Research Centre, University of Southampton, SO17 1BJ Southampton, UK*
Fazendorf@adva.com

**Abstract:** Using a Correlation-OTDR, we characterized the temperature-induced group delay variations of two nested antiresonant nodeless hollow core fibers. The temperature sensitivity of both is substantially less than for SSMF with some dependency on coating type. © 2021 The Author(s)

## 1. Introduction

The group delay of an optical signal in the fiber becomes a critical parameter for 5G networks and beyond. For some applications, like long single span transmission, the absolute delay value is of concern. For other applications, the differential delay between two optical paths is critical. One of these applications is the synchronization between a master and a slave clock. In current synchronization applications, like IEEE 1588v2 (PTP), an asymmetry of a few nanoseconds is tolerable. Future synchronization applications will likely require an asymmetry reduced by an order of magnitude. The group delay of the optical signal in a fiber is affected by temperature variations in the fiber environment. A typical value for the temperature coefficient of delay (TCD) of a standard single mode fiber (SSMF) is 7.49 ppm/K [1]. Consequently, the group delay of a 100 km fiber link changes by 3.75 ns per Kelvin. The TCD value comprises the thermal expansion of the fiber and the temperature induced change of the refractive index. One of the lowest-TCD fiber is a hollow core fiber (HCF), which consists of an air-filled core and a microstructured glass cladding [2]. As the refractive index of air changes less with temperature than fused silica, the TCD of HCF is significantly smaller than that of SSMF, which has been shown in other publications [2, 3]. It was reported that the protective fiber coating, comprising materials with different elastic moduli and coefficients of thermal elongation to silica, can also contribute to larger group delay changes [4]. In this paper, we use correlation-optical time domain reflectometry (C-OTDR) for precise measurements of the group delay variations of two nested antiresonant nodeless hollow core fibers (NANF-HCFs) with different coatings. We show that the group delay of an HCF changes with temperature and compare the variation between the different coatings.

## 2. Measurement setup

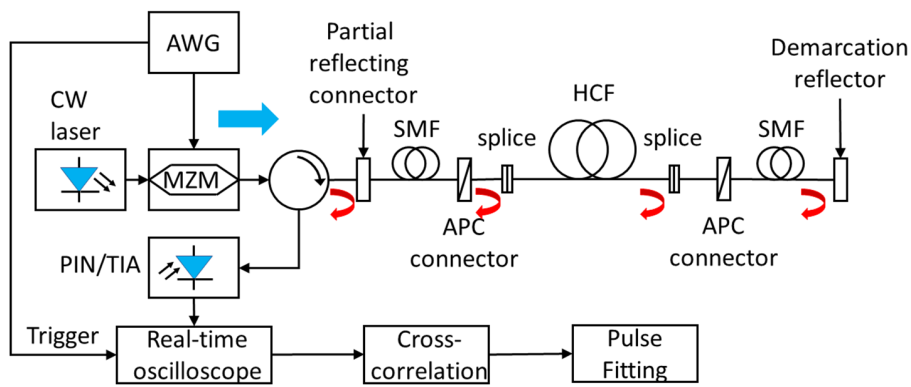

Fig. 1. Measurement setup of the C-OTDR to determine the group delay of two NANF- HCF.

The difference between the typical OTDR and the C-OTDR is that, for the latter, bursts of bits are sent instead of pulses. The typical OTDR is limited in time resolution by the probe pulse width. The time resolution of the C-OTDR was increased by sending a burst with a data rate of 10 Gbit/s into the fiber under test (FuT) and correlating the reflected signal with the transmitted burst. Through the correlation we obtain a time resolution of 100 ps, which corresponds to the bit rate of the probe signal. Additionally, by further signal processing steps like oversampling and pulse approximation, the precision of this measurement method was increased to a few picoseconds [5].

In Fig. 1 the schematic of the C-OTDR is illustrated. The blue arrow in Fig. 1 indicates the burst sent into the fibers under test. An arbitrary waveform generator (AWG) is used to generate the electrical signal, which consists of a 10-Gbit/s, 128-bit burst followed by zeros. The zeros, whose number depends on the fiber length, are essential to obtain unambiguous reflections. The AWG drives a Mach-Zehnder modulator, which modulates the generated sequence on a continuous wave signal at 1550 nm. The probe signal was sent into the optical circulator and was reflected first at the partial reflecting connector (red arrow) to obtain a reference. The transmitted part of the probe signal propagated trough the SMF jumper and was reflected at the first splice, which constitutes a transition between SMF and HCF and thus a change of the refractive index from glass to air. Further reflections were observed at the HCF-SMF connection and, after a second SMF jumper, at the broadband demarcation reflector. All reflected signals were received by a PIN/TIA combination and recorded by a real-time oscilloscope with a sampling rate of 50 GS/s. 1000 recorded traces were averaged and processed offline. The obtained signals were cross correlated with the transmitted burst and, after probing with two complementary Golay sequences, the correlation functions were added. The obtained correlation sum of one FuT is shown in Fig. 2. The sample values in the correlation sum have a spacing of 20 ps. To achieve a timing resolution better than one sample period, a Gaussian function was fitted to the correlation peak resulting from each reflection. An example of the Gaussian fit to the second correlation peak of the trace is embedded in Fig. 2. The center of the Gaussian function is recorded as the precise round-trip group delay to the reflection location.

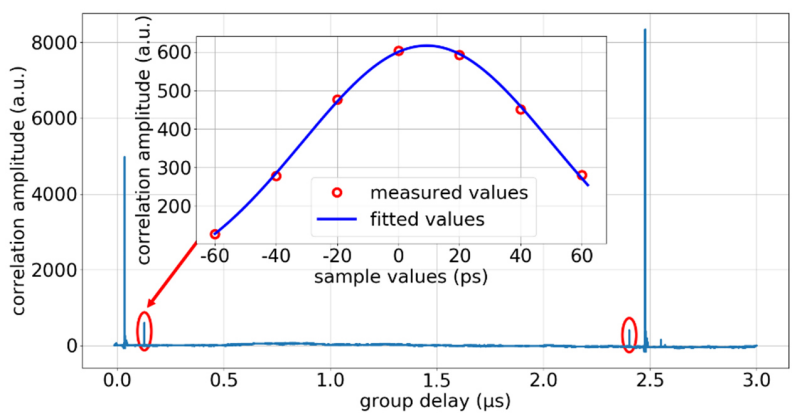

Fig. 2. Correlation function over time shows four reflections: two introduced reflections and two reflections from glass to air transition. Inset: pulse approximation of one reflection peak.

## 3. Results

In the correlation function in Fig. 2, the reflections from the glass air transition are marked with red ellipses. They allow to measure the absolute group delay of the FuT without having to take into account the length of the pigtails. Two HCF samples were used: a 340 m long HCF with a double coating, which means that two layers of coating with different material properties were used, similarly to SSMF. The other HCF was 289 m long and had only one layer of coating. Using a refractive index of n=1, the expected round-trip group delays were 2268.235 ns for the 340-m HCF and 1928.004 ns for the 289-m HCF, respectively. The measured group delays of 2275.099 ns and 1938.809 ns for the two HCFs point to a slightly varying effective group index of 1.003 and 1.005, respectively. To analyze the group delay variations of both HCF, the fiber samples were placed in a temperature-controlled cabinet. The temperature was varied in the range of 10 °C to 50 °C. After a settlement time of 1 h, the group delay was measured ten times at each temperature using the C-OTDR method. In Fig. 3a, the group delay variations over temperature are shown for both fibers. The measured values are shown as dots, and for each series the average value was calculated showing as straight line with dots as markers for the double coated HCF and rectangle as markers for the single coated HCF. The values measured for the double coated HCF have a standard deviation of 0.5 ps for each temperature point. The measured values for the single coated HCF have a higher standard deviation of 1.5 ps. This is likely caused by the lower sampling rate and the lower voltage resolution of the oscilloscope used in the measurement of the single coated HCF. We observed a nonlinear behavior of the group delay vs. temperature for both HCFs. This might be caused by the temperature dependence of the material properties of the coatings. We observed a similar behavior in jumper cables with a tight buffer jacket [6]. In Fig. 3a, the dashed line with rectangular markers shows a TCD of 1.5 ppm/K as the average for the double coated HCF. This value leads to a round-trip group delay change of 3.42 ps/K over a length of

340 m. For the single-coated HCF, we measured a TCD of 0.55 ppm/K shown by the dashed line with triangle markers in the temperature range from 10°C to 35°C. Our measurement result fits with previously published works [2,4]. However, the group delay variations change for higher temperatures, observed as a bounce of the group delay at 40°C. This bounce might be caused by tight spooling of the fiber, which leads to more stress on the fiber due to the thermal expansion of the plastic spool (15.5cm diameter. This behavior was repeatable over four temperature cycles. The measured TCD of 0.55 ppm/K leads to a round-trip group delay change of 1.06 ps/K, while the measured TCD of 0.94 ppm/K leads to a group delay change of 1.82 ps/K for 282 m of HCF.

For a sweep with decreasing temperature from 50°C to 10°C, the drop in group delay was already observed above 40°C such that we obtained an inverse hysteresis, as illustrated in Fig. 3b. A similar inverse hysteresis was observed in a previous work with SMF jumpers [6]. Nevertheless, it is noticeable that the coating influences the temperature dependence of the group delay.

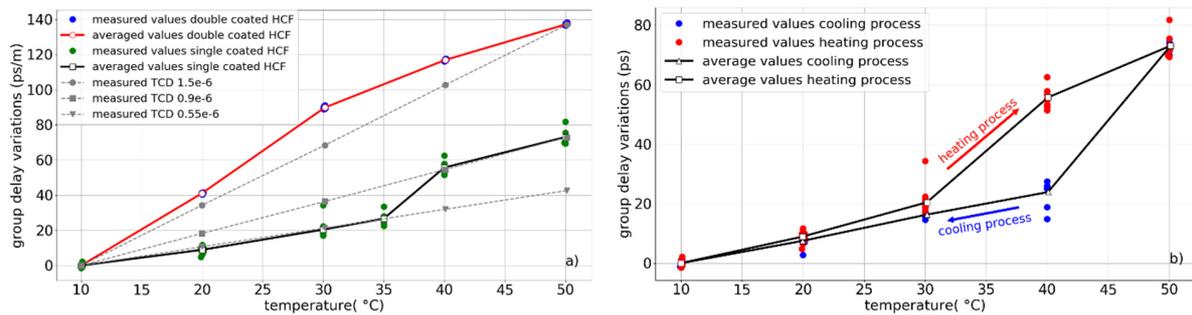

Fig. 3 a) Measured group delay variations as function of temperature with measured TCD as linear fit, b) Measured group delay variations of the single coated HCF during heating and cooling process which shows a hysteresis.

## 4. Conclusion

Hollow core fiber has a 5 to 13 times lower TCD than SSMF making it a candidate for synchronization applications where a stable and symmetric group delay is required. Using a Correlation OTDR with an accuracy of approximately 3 ps, we were able to characterize the temperature dependent group delay of two HCF samples with different coatings. The double-coated sample displays a two times higher temperature dependence of the group delay as compared to the single coated one. We attribute this to the difference between the moduli and thermal expansion coefficients of the various coating materials employed. Given the differences observed between these single and double-coated HCFs (which use conventional coating recipes) we believe there to be considerable scope to engineer the coating design to provide even lower values of TCD.

## 5. Acknowledgements

This work was funded by the German Federal Ministry of Education and Research (BMBF) under the project OptiCON with grant agreement number 16KIS0989K and UK EPSRC project Airguide.

**References**

[1] G. Lutes and W. Diener, "Thermal coefficient of delay for various coaxial and fiber-optic cables," TDA Progress Report 42-99, Nov. 1989, available online at https://ipnpr.jpl.nasa.gov/progress_report/42-99/99E.PDF
[2] W. Zhu et al., "Toward High Accuracy Positioning in 5G via Passive Synchronization of Base Stations Using Thermally-Insensitive Optical Fibers". IEEE Access Vol. 7, 2019
[3] Dangui, V., Kim, H. K., Digonnet, M. J. F. & Kino, G. S. "Phase sensitivity of temperature of the fundamental mode in air-guiding photonic-bandgap fibers". Opt. Express 13**,** 6669–6684 (2005).
[4] E. Numkam Fokoua, et al., "How to make the propagation time through an optical fiber fully insensitive to temperature variations", Optica Vol. 4, 659-668, 2017.
[5] F. Azendorf, et al., „Improvement of accuracy for measurement of 100-km fibre latency with Correlation OTDR ", ECOC 2019, Dublin, September 2019.
[6] F. Azendorf, A. Dochhan, and M. Eiselt, "Temperature Dependent Latency of Jumper Cable ", ITG Photonic Networks 2018, Leipzig, June 2018.